\documentclass[conference]{IEEEtran}
\IEEEoverridecommandlockouts
\usepackage{cite}
\usepackage{svg}
\usepackage{amsmath,amssymb,amsfonts}
\usepackage{algorithmic}
\usepackage{graphicx}
\usepackage{textcomp}
\usepackage{xcolor}
\usepackage{booktabs} 
\usepackage{caption}
\usepackage{lipsum} 
\usepackage{booktabs}

\usepackage{hyperref} 
\def\BibTeX{{\rm B\kern-.05em{\sc i\kern-.025em b}\kern-.08em
    T\kern-.1667em\lower.7ex\hbox{E}\kern-.125emX}}
    
\begin{document}

\title{Meta-Learning for Quantum Optimization via  Quantum Sequence Model

\thanks{The views expressed in this article are those of the authors and do not represent the views of Wells Fargo. This article is for informational purposes only. Nothing contained in this article should be construed as investment advice. Wells Fargo makes no express or implied warranties and expressly disclaims all legal, tax, and accounting implications related to this article.\\
\IEEEauthorrefmark{5} \href{mailto:an4114752@gs.ncku.edu.tw}{an4114752@gs.ncku.edu.tw}.
\IEEEauthorrefmark{6} \href{mailto:ycchen1989@ieee.org}{ycchen1989@ieee.org}.
}}

\author{
\IEEEauthorblockN{
    Yu-Cheng Lin \IEEEauthorrefmark{1},
    Yu-Chao Hsu\IEEEauthorrefmark{2}\IEEEauthorrefmark{3}\IEEEauthorrefmark{5},
    Samuel Yen-Chi Chen\IEEEauthorrefmark{4}\IEEEauthorrefmark{6},
}
\IEEEauthorblockA{\IEEEauthorrefmark{1} Department of Electrophysics, National Yang Ming Chiao Tung University, Hsinchu, Taiwan}
\IEEEauthorblockA{\IEEEauthorrefmark{2} National Center for High-Performance Computing, National Institutes of Applied Research, Hsinchu, Taiwan}
\IEEEauthorblockA{\IEEEauthorrefmark{3} Cross College Elite Program, National Cheng Kung University, Tainan, Taiwan}
\IEEEauthorblockA{\IEEEauthorrefmark{4} Wells Fargo, New York, NY, USA}
}

\maketitle

\begin{abstract}
The Quantum Approximate Optimization Algorithm (QAOA) is a leading approach for solving combinatorial optimization problems on near-term quantum processors. However, finding good variational parameters remains a significant challenge due to the non-convex energy landscape, often resulting in slow convergence and poor solution quality. In this work, we propose a quantum meta-learning framework that trains advanced quantum sequence models to generate effective parameter initialization policies. We investigate four classical or quantum sequence models, including the Quantum Kernel-based Long Short-Term Memory (QK-LSTM), as learned optimizers in a "learning to learn" paradigm. Our numerical experiments on the Max-Cut problem demonstrate that the QK-LSTM optimizer achieves superior performance, obtaining the highest approximation ratios and exhibiting the fastest convergence rate across all tested problem sizes ($n=10$ to $13$). Crucially, the QK-LSTM model achieves perfect parameter transferability by synthesizing a single, fixed set of near-optimal parameters, leading to a remarkable sustained acceleration of convergence even when generalizing to larger problems. This capability, enabled by the compact and expressive power of the quantum kernel architecture, underscores its effectiveness. The QK-LSTM, with only 43 trainable parameters, substantially outperforms the classical LSTM (56 parameters) and other quantum sequence models, establishing a robust pathway toward highly efficient parameter initialization for variational quantum algorithms in the NISQ era.
\end{abstract}


\section{Introduction}
Optimization problems are widely applied across fields such as finance\cite{naik2025portfolio}, logistics\cite{rahmanifar2025green}, and healthcare\cite{cabrera2012simulation}. Despite significant progress, classical optimization algorithms face limitations in computational resources and algorithmic constraints when dealing with large-scale or complex problems, preventing efficient handling.   

To overcome these challenges, quantum optimization methods, particularly the Quantum Approximate Optimization Algorithm (QAOA)~\cite{farhi2014quantum} is proposed to solve combinatorial optimization problems and other NP-hard tasks~\cite{chatterjee2024solving} using shallow quantum circuits on near-term quantum devices\cite{zhou2020quantum,lotshaw2022scaling,chen2024noise}.

Despite its potential advantages, optimizing the parameters of QAOA remains a significant challenge. The optimization landscape is inherently non-convex, characterized by multiple local minimal \cite{chen2025L2L}, while the performance is further affected by hardware noise and measurement uncertainties present in near-term quantum devices. From the viewpoint of classical optimization, Bittel et al.~\cite{bittel2021training} showed that, even if a quantum system can be efficiently simulated classically, the corresponding classical optimization process still remains NP-hard.

Moreover, the presence of barren plateaus\cite{mcclean2018barren,holmes2022connecting,wang2021noise} and narrow gorges\cite{arrasmith2022equivalence} during training significantly complicates the optimization process, often preventing the model from achieving stable convergence. To address this issue, meta-learning\cite{hospedales2021meta,lee2025q} has emerged as a promising paradigm that enables models to learn how to optimize by leveraging prior experience across multiple related tasks.

The concept of a learned optimizer was first introduced by Andrychowicz et al.~\cite{andrychowicz2016learning}, where a coordinate-wise Long Short-Term Memory (LSTM) network was used to replace traditional hand-crafted optimizers.

In this work, we propose a quantum sequence model that integrates meta-learning with quantum-enhanced learning-based optimizers to substantially improve the efficiency and performance of quantum optimization algorithms.
Specifically, we develop three meta-learned architectures within this framework such as Quantum Long Short-Term Memory (QLSTM)\cite{chen2022quantum,chen2025toward}, Quantum Kernel-based LSTM (QK-LSTM)\cite{hsuLSTM2025quantum,hsu2025quantum,hsu2025federated}, and Quantum Fast Weight Programmer (QFWP)\cite{chen2024learning,chen2025learningProgram,liu2025programming}, each designed to capture different aspects of temporal dynamics and optimization behavior in quantum systems. We show that in Section \ref{sec:Quantum Sequence Model}.

In this way our approach demonstrates significant improvements in convergence speed and solution quality.
In particular, QK-LSTM and QLSTM outperform conventional classical optimization methods across multiple benchmark tests.
Furthermore, we conduct an extensive benchmarking study, systematically comparing these quantum meta-learning architectures with various classical optimizers to comprehensively evaluate their performance and robustness.
The experimental results are presented and discussed in Section~\ref{result}. Finally, we discuss and conclude the paper in Section~\ref{sec:dis} and ~\ref{sec:con}.

\section{Quantum Sequence Model}\label{sec:Quantum Sequence Model}
\subsection{Classical Long Short-Term Memory}
\begin{figure}[h]
     \centering
    \includegraphics[width=0.48\textwidth]{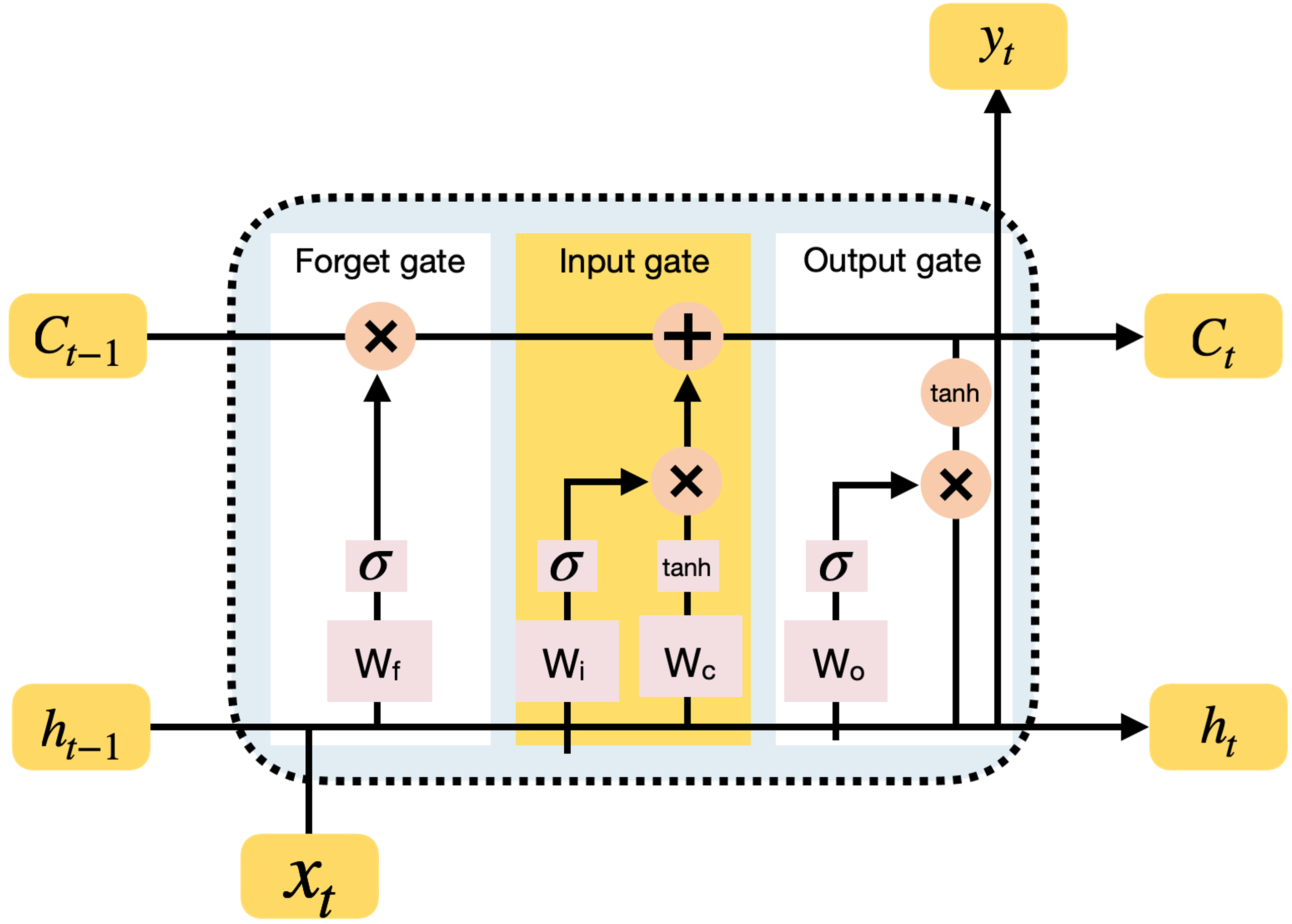}
    \caption{
     Schematic representation of the LSTM.
    }
    \label{LSTM}
    \vspace{-12pt}
\end{figure}

The LSTM networks
(see Fig. \ref{LSTM}) have profoundly influenced the development of machine learning and sequential data modeling \cite{hochreiter1997long}.
By effectively addressing the vanishing and exploding gradient problems that limit the Recurrent Neural Network (RNN)\cite{medsker2001recurrent}, it offers a stable mechanism for learning long-term dependencies in sequential data. LSTM networks have been widely applied in various fields, including time series 
forecasting\cite{siami2018forecasting,srivastava2022weather,shi2018lstm}, natural language processing\cite{yao2018improved,wang2015learning,lippi2019natural,azari2019energy}, and medical applications\cite{lipton2015learning,edara2023sentiment,choppara2025leveraging}.
In this work, we employ LSTM networks for meta-learning\cite{younger2001meta}.

\begin{subequations}
\allowdisplaybreaks
\vspace{-3pt}
\begin{align}
\mathbf{f}_t &= \sigma(\mathbf{W}_f [\mathbf{h}_{t-1}, \mathbf{x}_t] + \mathbf{b}_f), \\
\mathbf{i}_t &= \sigma(\mathbf{W}_i [\mathbf{h}_{t-1}, \mathbf{x}_t] + \mathbf{b}_i), \\
\tilde{\mathbf{c}}_t &= \tanh(\mathbf{W}_c [\mathbf{h}_{t-1}, \mathbf{x}_t] + \mathbf{b}_c), \\
\mathbf{c}_t &= \mathbf{f}_t \odot \mathbf{c}_{t-1} + \mathbf{i}_t \odot \tilde{\mathbf{c}}_t, \\
\mathbf{o}_t &= \sigma(\mathbf{W}_o [\mathbf{h}_{t-1}, \mathbf{x}_t] + \mathbf{b}_o), \\
\mathbf{h}_t &= \mathbf{o}_t \odot \tanh(\mathbf{c}_t)
\end{align}
\vspace{-3pt}
\end{subequations}

where:
\begin{itemize}
    \item \( x_t \in \mathbb{R}^n \) is the input vector at time \( t \),
    \item \( h_{t-1} \in \mathbb{R}^m \) is the hidden state from the previous time step,
    \item \( W_f, W_i, W_C, W_o \) are weight matrices,
    \item \( b_f, b_i, b_C, b_o \) are bias vectors,
    \item \( \sigma(\cdot) \) denotes the sigmoid activation function,
    \item \( \tanh(\cdot) \) denotes the hyperbolic tangent activation function,
    \item \( \odot \) represents element-wise multiplication.
\end{itemize}

\subsection{Quantum Long Short-Term Memory}
\begin{figure}[h]
     \centering
    \includegraphics[width=0.48\textwidth]{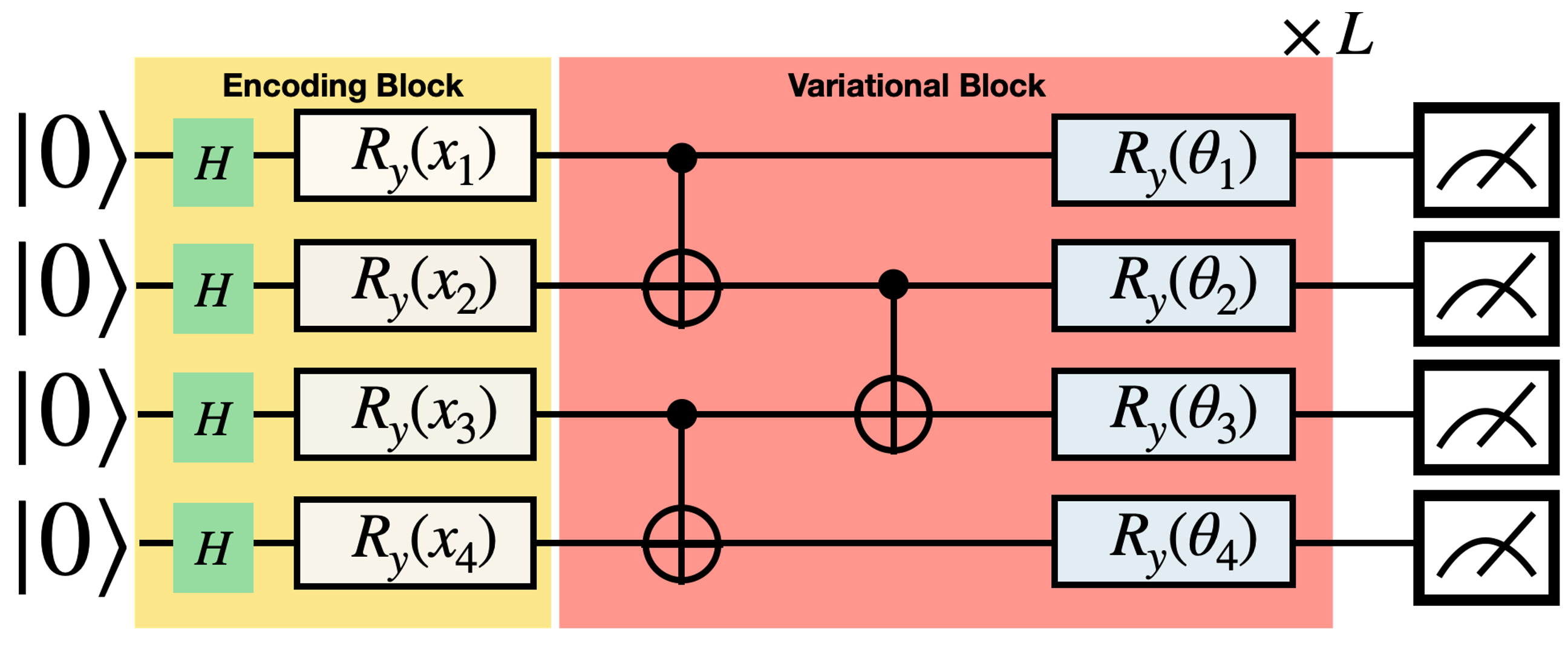}
    \caption{
     Schematic representation of the VQC.
    }
    \label{VQC}
    \vspace{-12pt}
\end{figure}

\begin{figure}[h]
     \centering
    \includegraphics[width=0.48\textwidth]{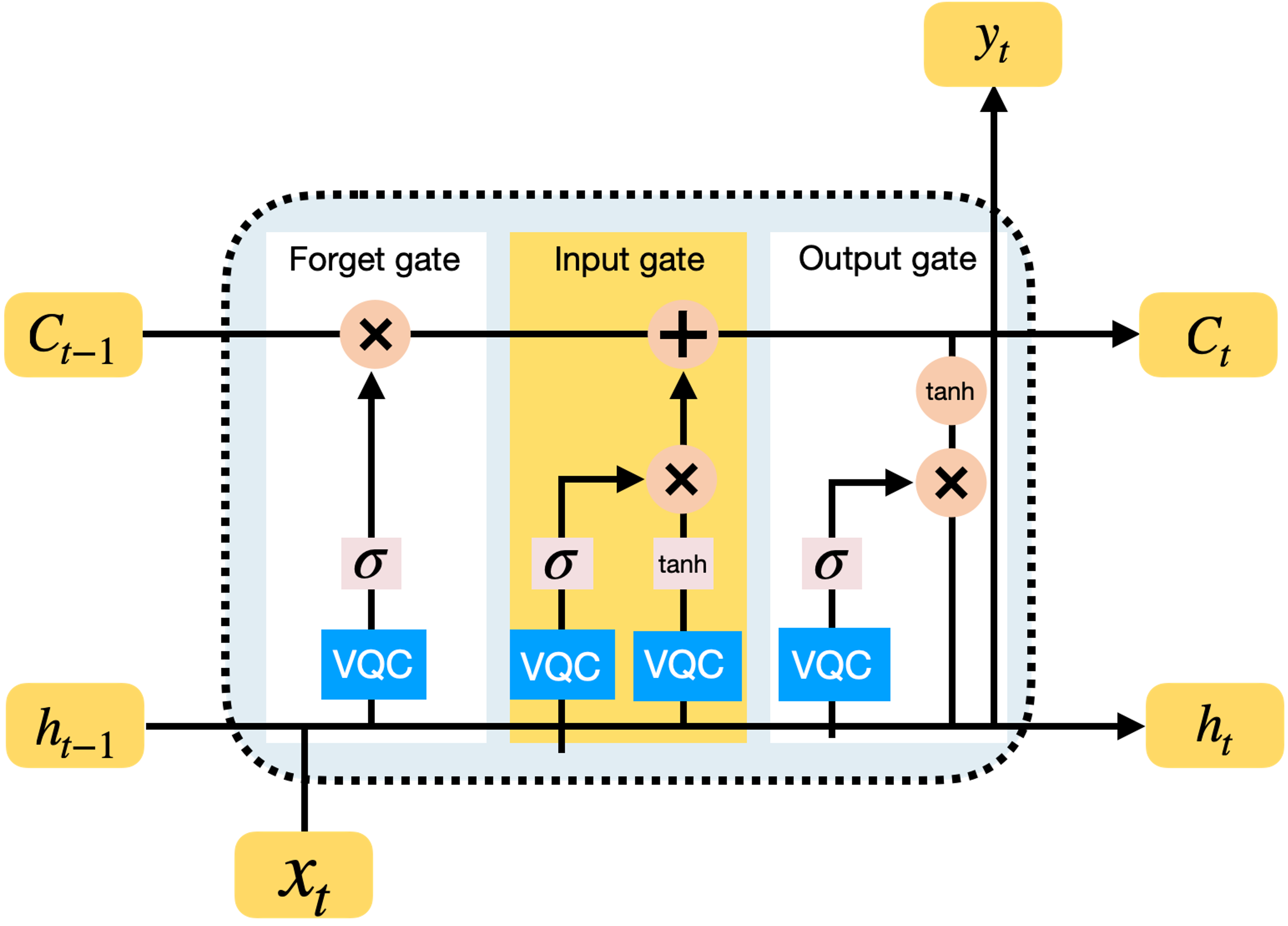}
    \caption{
     Schematic representation of the QLSTM.
    }
    \label{QLSTM}
    \vspace{-12pt}
\end{figure}

The QLSTM model\cite{chen2022quantum} extends the classical LSTM by replacing its classical neural network components with variational quantum circuit (VQC)\cite{chen2020variational,khairy2020learning,hsu2025Adaptive}, thereby exploiting quantum mechanical properties such as superposition and entanglement (see Fig. \ref{VQC} and Fig. \ref{QLSTM}). The core operation of the QLSTM lies in its gating mechanism, which regulates the flow of information through quantum-encoded states.

\begin{subequations}
\allowdisplaybreaks
\vspace{-3pt}
\begin{align}
\mathbf{f}_t &= \sigma(\text{VQC}_1(v_t)), \\
\mathbf{i}_t &= \sigma(\text{VQC}_2(v_t)), \\
\tilde{\mathbf{c}}_t &= \tanh(\text{VQC}_3(v_t)), \\
\mathbf{c}_t &= \mathbf{f}_t \odot \mathbf{c}_{t-1} + \mathbf{i}_t \odot \tilde{\mathbf{c}}_t, \\
\mathbf{o}_t &= \sigma(\text{VQC}_4(v_t)), \\
\mathbf{h}_t &= \mathbf{o}_t\odot \tanh{(\mathbf{c}_t}),\\
\end{align}
\vspace{-3pt}
\end{subequations}
where \( v_t = [h_{t-1}, x_t] \) denotes the concatenation of the current input \( x_t \) at time step \( t \) and the hidden state \( h_{t-1} \) from the previous time step.

\subsection{Quantum Kernel-Based Long Short-Term Memory}
\begin{figure}[h]
     \centering
    \includegraphics[width=0.48\textwidth]{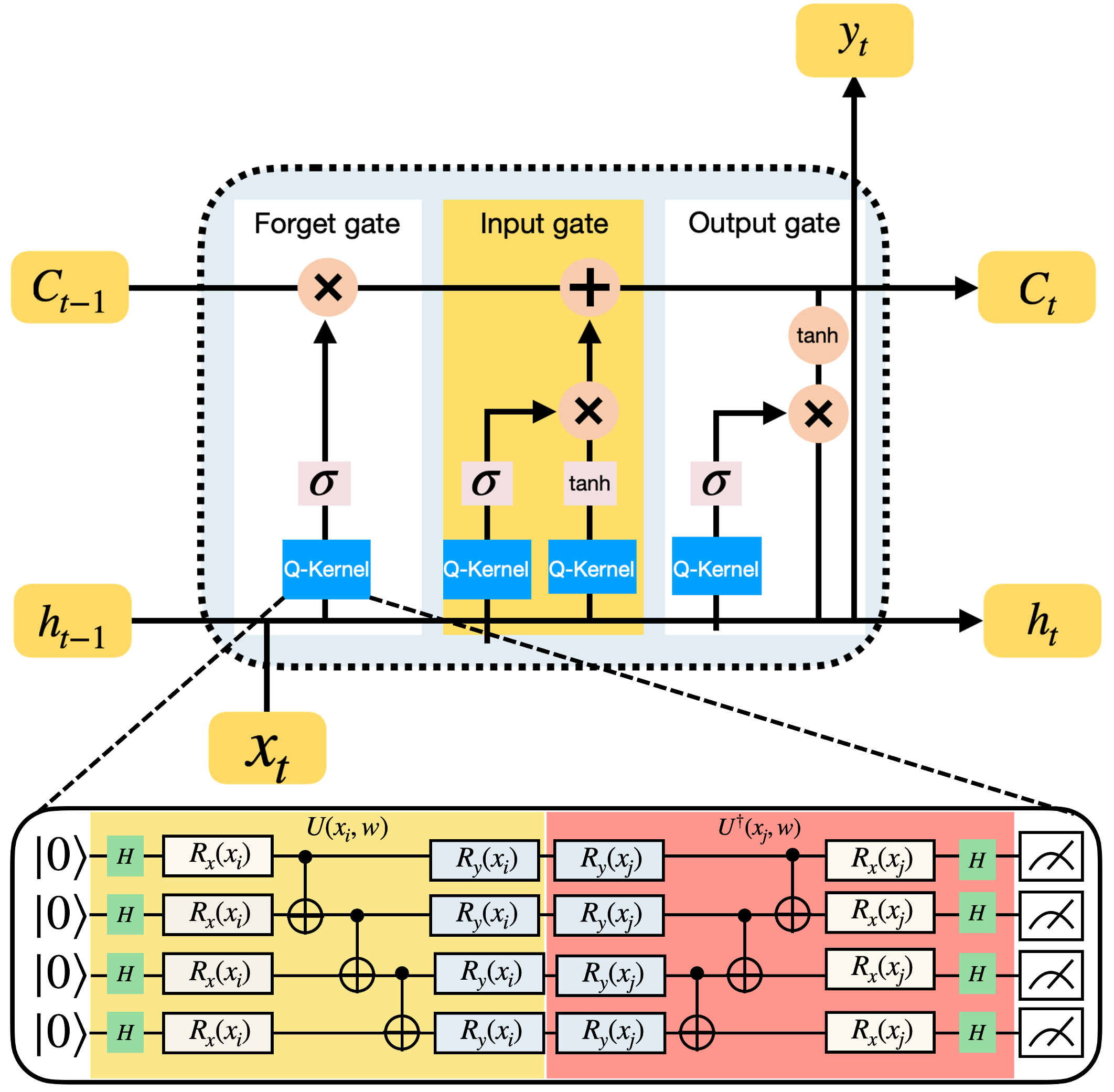}
    \caption{Schematic illustration of the QK-LSTM and the quantum circuit representation of \(U(x_i, w)\) and \(U^{\dagger}(x_j, w)\), demonstrating the use of quantum gates for encoding data and extracting features for machine learning applications.}

    \label{QK-LSTM}
    \vspace{-12pt}
\end{figure}

The QK-LSTM Network\cite{hsuLSTM2025quantum,hsu2025quantum,hsu2025federated}, a hybrid machine learning architecture that combines the strengths of classical LSTM networks with quantum kernel method\cite{chen2024validating,thanasilp2024exponential,tsai2025learning}  (see Fig. \ref{QK-LSTM}). This model is employed to enhance performance and reduce the number of trainable parameters in meta-learning tasks.

The formal mathematical formulation of the QK-LSTM cell is defined as follows:

\begin{subequations}
\allowdisplaybreaks
\vspace{-3pt}
\begin{align}
f_t &= \sigma\left( \sum_{j=1}^{N} \beta_j^{(f)} \, \kappa^{(f)}(v_t, v_j) \right), \\
i_t &= \sigma\left( \sum_{j=1}^{N} \beta_j^{(i)} \, \kappa^{(i)}(v_t, v_j) \right), \\
\hat{C}_t &= \tanh\left( \sum_{j=1}^{N} \beta_j^{(C)} \, \kappa^{(C)}(v_t, v_j) \right), \\
C_t &= f_t \odot C_{t-1} + i_t \odot \hat{C}_t, \\
o_t &= \sigma\left( \sum_{j=1}^{N} \beta_j^{(o)} \, \kappa^{(o)}(v_t, v_j) \right), \\
h_t &= o_t \odot \tanh\left( C_t \right),
\end{align}
\vspace{-3pt}
\end{subequations}
\noindent
where:
\begin{itemize}
    \vspace{+3pt}
    \item  \(v_t = [h_{t-1}; x_t] \in \mathbb{R}^{n + m}\) represents the concatenation of input \(x_t\) at time-step \(t\) and the hidden state \(h_{t-1}\) from the previous time-step \(t-1\),
    \vspace{+3pt}
    \item \( \beta_j^{(f)} \), \( \beta_j^{(i)} \), \( \beta_j^{(C)} \), and \( \beta_j^{(o)} \) are trainable coefficients corresponding to each gate's quantum kernel,
    \vspace{+3pt}
    \item \( \kappa^{(f)}(\cdot, \cdot) \), \( \kappa^{(i)}(\cdot, \cdot) \), \( \kappa^{(C)}(\cdot, \cdot) \), and \( \kappa^{(o)}(\cdot, \cdot) \) denote the quantum kernel functions tailored to the respective gates.
    
\end{itemize}

\subsection{Quantum FWP}
The classical Fast Weight Programmer (FWP) model~\cite{schmidhuber1992learning} consists of two neural networks: a \textit{slow network} and a \textit{fast network}, where the weights of the fast network act as the program. 
The slow network updates the weights of the fast network at each time step based on incoming data, without completely overwriting them. 
This mechanism enables the model to perform rapid short-term adaptation while preserving long-term information.In this way, it eliminates the need for recurrent connections, thereby reducing computational overhead.

\begin{figure}[h]
     \centering
    \includegraphics[width=0.48\textwidth]{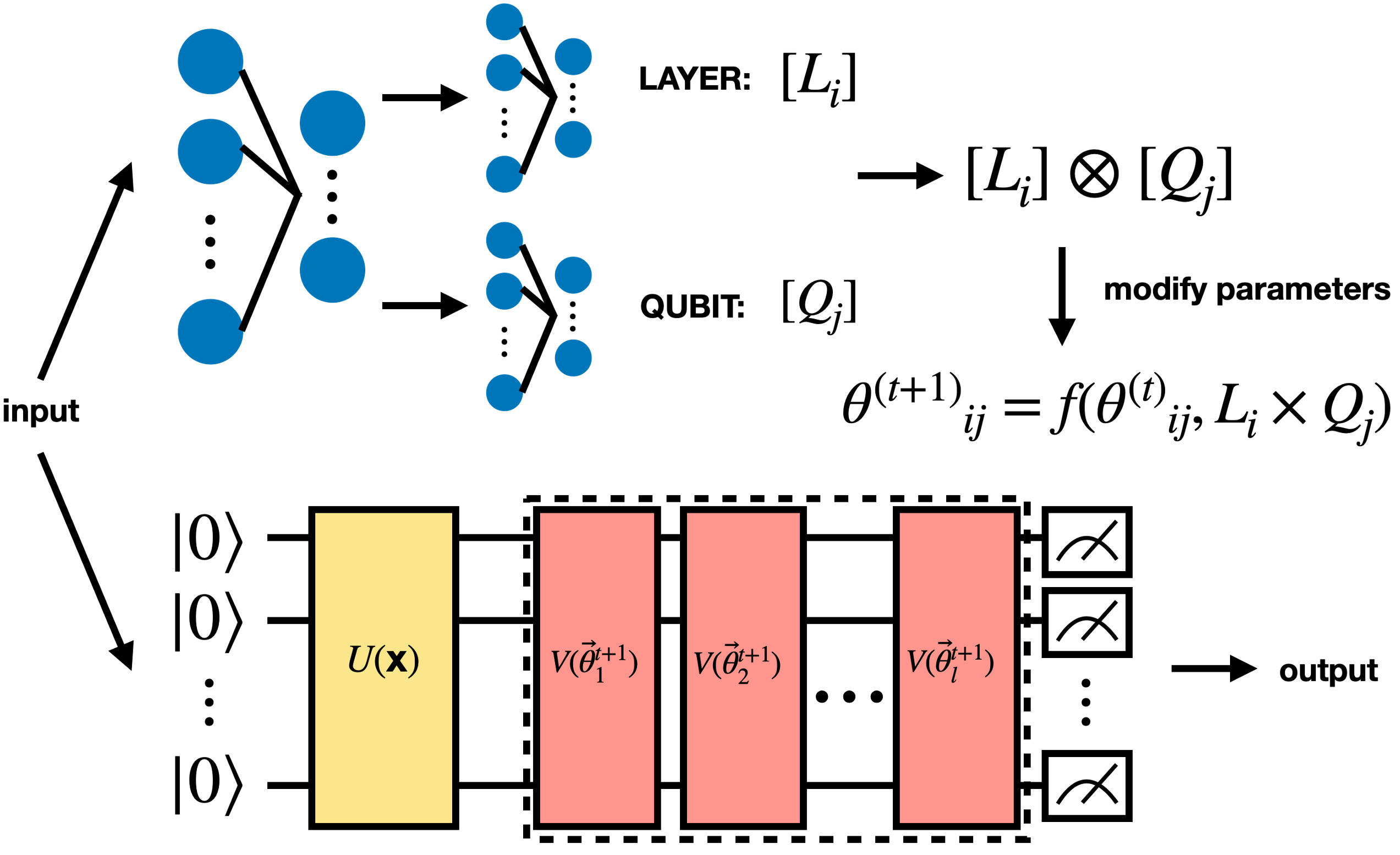}
    \caption{
     Schematic representation of the QFWP.
    }
    \label{QFWP}
    \vspace{-12pt}
\end{figure}

Inspired by this architecture, the Quantum Fast Weight Programmer
(QFWP) (see Fig. \ref{QFWP}) \cite{chen2024learning} replaces the classical neural networks of both the slow and fast programs with VQC. 
The classical input vector $\vec{x}$ is mapped to latent vectors $\mathbf{L} \in \mathbb{R}^L$ and $\mathbf{Q} \in \mathbb{R}^n$.  

The outer product $\mathbf{L} \times \mathbf{Q}$ is then computed to generate an update term for the VQC parameters. At each time step $t+1$, the parameters of the quantum circuit are updated according to
$\theta_{ij}^{t+1} = f(\theta_{ij}^t, \mathbf{L}_i \times \mathbf{Q}_j)$,
where $f(\cdot)$ combines the parameters from the previous time step $\theta_{ij}^t$ with the newly generated term $\mathbf{L}_i \times \mathbf{Q}_j$. This approach enables the circuit parameters to retain information from previous time steps. The output of the VQC can be further refined through post-processing layers, such as normalization, translation, or classical neural networks, to enhance the final results.

\begin{figure}[b]
     \centering
    \includegraphics[width=0.48\textwidth]{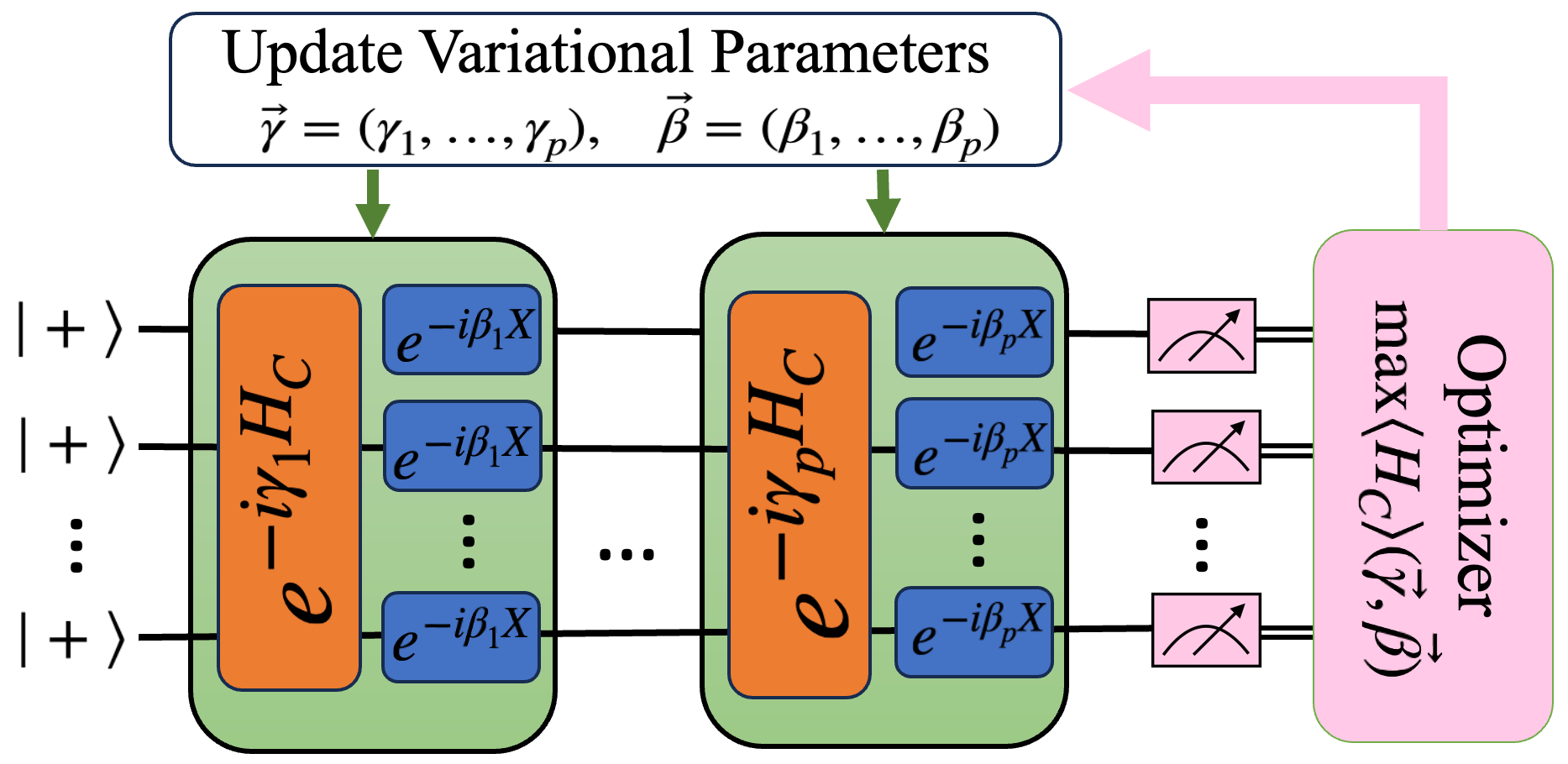}
    \caption{
     Schematic representation of the QAOA.
    }
    \label{QAOA}
    \vspace{-12pt}
\end{figure}

\section{Method}\label{sec:met}

All numerical simulations and model training processes were performed using Python (v3.10.18).The primary libraries utilized include PennyLane (v0.42.3) \cite{antipov2022pennylane} for quantum circuit simulation, PyTorch (v2.8.0) \cite{2019torch} for neural network implementation and training, and SciPy (v1.15.3) \cite{roy2023scipy} for numerical operation. All experiments were executed on a system equipped with eight NVIDIA Tesla V100 GPUs (16GB) and Intel Xeon CPUs.

\subsection{Quantum Approximate Optimization Algorithms for MaxCut}

This work addresses the optimization challenge of the QAOA (see Fig. \ref{QAOA}) applied to the Max-Cut problem on an unweighted graph $G\in (V,E)$\cite{farhi2014quantum}. The max-cut problem is formulated as identifying a vertex partition that maximizes the number of edges crossing the cut.

The QAOA ansatz acts on the initial uniform superposition state $|\psi_0\rangle=|+\rangle^{\otimes n}$, where $n=|V|$ is the number of qubits. The variational state is then defined by the alternating application of the cost and mixer Hamiltonians: 

\allowdisplaybreaks
\vspace{-3pt}
\begin{align}
|\psi(\vec{\theta})\rangle = \prod_{j=1}^P e^{-i\beta_j \hat{H}_M}e^{-i\gamma_j\hat{H}_C}|\psi_0\rangle
\end{align}
\vspace{-3pt}

where $\vec{\theta}=(\vec{\beta}, \vec{\gamma}) \in \mathbb{R}^{2P}$, and $\vec{\beta}$ and $\vec{\gamma}$ is the vector of variational parameters, and $P$ is the number of layers and the cost Hamiltonian $\hat{H}_C$ encodes the Max-Cut objective via a sum of pairwise $\hat{Z}$ operators:

\allowdisplaybreaks
\vspace{-3pt}
\begin{align}
\hat{H}_C=\sum_{\{j,k\}\in e}\frac{1}{2}(\hat{I}-\hat{Z}_j\hat{Z}_k)
\end{align}
\vspace{-3pt} 

In this study, the QAOA layer is fixed at $P=1$, yielding a total of $2$ variational parameters, $\vec{\theta}=(\beta_1, \gamma_1)$. 

The objective function to be maximized is the expected cut value:

\allowdisplaybreaks
\vspace{-3pt}
\begin{align}
f(\vec{\theta})=\langle \psi(\vec{\beta},\vec{\gamma})|\hat{H}_C|\psi(\vec{\beta},\vec{\gamma})\rangle
\end{align}
\vspace{-3pt} 

We generate random problem instances using the Erdős–Rényi graph model. For training, we sampled $1008$ graphs with $n\in [6,9]$ nodes. For testing, $90$ graphs with $n\in [10,13]$ nodes are generated. In both cases, the connectivity probability of each edge is $p=k/n$, where $k\in [3, n-1]$.

\subsection{Meta-Learning Framework for QAOA}
The main challenge in optimizing the QAOA lies in the inherent inefficiency and susceptibility to local minima within the classical optimization loop. To overcome this, we use the parameter tuning process as a sequence prediction problem within the meta-learning paradigm, commonly referred to as "learning to learn" (L2L) \cite{andrychowicz2016learning}.

\begin{figure}[h]
    \centering
    \includegraphics[width=0.48\textwidth]{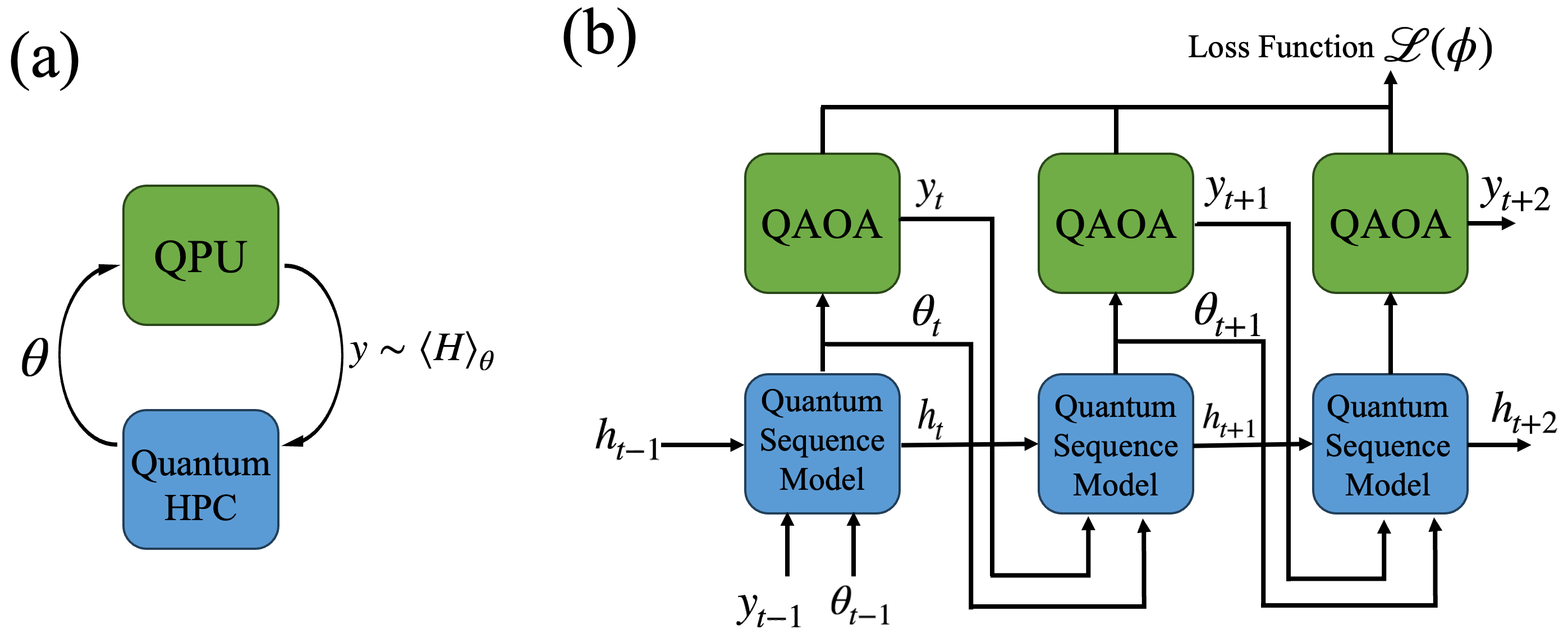}
    \caption{
        Schematic representation of the QAOA framework with Meta Learning. (a) Quantum processing unit (QPU) interacting with Quantum HPC, where the quantum state \(\theta\) is evaluated with the Hamiltonian \(\langle H \rangle_\theta\). (b) Architecture of a quantum sequence model using QAOA blocks for optimizing time-series predictions.
    }
    \label{L2L}
    \vspace{+10pt}
\end{figure}

Taking inspiration from the pioneering work of Verdon \textit{et al.} \cite{Guillaume2019lstmvqe}, which utilized RNN for quantum parameter initialization (see Fig. \ref{L2L}), we extend their approach to a comparative study encompassing a suite of advanced quantum and classical sequence models: QK-LSTM, QLSTM, and QFWP. In this meta-learning context, the objective is to train an optimization agent ($\text{Model}_{\phi}$) to generate an effective update policy for the QAOA variational parameters ($\vec{\theta}$) based on the historical trajectory of the optimization. The underlying intuition for this approach is that the variational parameters ($\gamma$ and $\beta$) are analogous to hyperparameters governing descent dynamics in the rugged cost landscape \cite{farhi2014quantum}. A sequence model, trained on diverse small problem instances, learns a generalized, problem-agnostic update that effectively approximates the near-optimal control sequence, thereby significantly reducing the quantum-classical iterations required for initialization. 

For consistency and fair comparison across the models investigated, structural adjustments were implemented to match the output dimension of the QAOA parameters. Four qubits were used for the internal quantum operations within the QLSTM and QK-LSTM models. Since the QAOA parameter dimension of $4$, a fully connected (FC) layer was appended downstream to the QK-LSTM and QLSTM models to reshape their output to the required parameter size, while the standard LSTM and QFWP models were structured to output the final parameter size directly. 

\subsection{Optimization Protocol}
Our optimization strategy consists of a meta-learning process to train the sequence models, followed by a two-phase optimization protocol during testing to evaluate performance.

\subsubsection{Model Training (Meta-Learning)}
The models were trained using small graphs $n\in[6,9]$ to learn an effective parameter update policy $\phi$. The training process utilized a fixed time horizon of $T=10$ steps. The model functions as a meta-optimizer, which use the previous state $h_t$, the parameter vector $\vec{\theta}_t$, and the estimated cost $y_t$ at each step $t$ to predict the next parameter set:

\allowdisplaybreaks
\vspace{-5pt}
\begin{align}
h_{t+1}, \vec{\theta}_{t+1} = \text{Model}_{\phi}(h_t, \vec{\theta}_t, y_t)
\end{align}
\vspace{-5pt}

The initial parameters for the training loop were set to $\vec{\theta}_0 = (0^{\otimes 2P})$. The model parameters $\phi$ were optimized using RMSprop via backpropagation through time \cite{Pascanu2013BTT} for a maximum of 50 epochs.

The meta-loss $\mathcal{L}(\phi)$ was computed as a weighted sum of the QAOA costs $f(\vec{\theta})$ across 10 steps:

\allowdisplaybreaks
\vspace{-3pt}
\begin{align}
\mathcal{L}(\phi)=\mathbb{E}[\sum_{t=1}^T 0.1t\cdot f(\vec{\theta_t})]
\end{align}
\vspace{-3pt}

All models were trained with identical sequence model hyperparameters for rigorous comparison, as detailed in Table \ref{tab:parameter}.

\begin{table}[t]
\caption{COMPANISON OF PARAMETERS FOR SEQUENCE MODELS}
\centering
\label{tab:parameter}
\resizebox{\columnwidth}{!}{%
\begin{tabular}{lcccc}
\toprule
\textbf{Parameter} & \textbf{QK-LSTM} & \textbf{LSTM} & \textbf{QLSTM} & \textbf{QFWP} \\
\midrule
Epochs & 50 & 50 & 50 & 50 \\
Recurrent Steps ($T$) & 10 & 10 & 10 & 10 \\
Learning Rate (Sequence Model) & $6\times 10^{-6}$ & $6\times 10^{-6}$ & $6\times 10^{-6}$ & $6\times 10^{-6}$ \\
Learning Rate (FC Layer) & $1\times 10^{-4}$ & - & $1\times 10^{-4}$ & - \\
Number of Qubits in Circuit & 4 & - & 4 & 2 \\
\midrule
\textbf{Total Trainable Parameters} & \textbf{43} & \textbf{56} & \textbf{43} & \textbf{31}\\
\bottomrule
\end{tabular}%
}
\end{table}

\subsubsection{Test Optimization Protocol}
For testing on larger, unseen graphs $n\in[10,13]$, a two-phase optimization protocol was executed. In the first phase (Phase I), the trained sequence model was run for $10$ steps starting from $\vec{\theta}_0$ to generate a high-quality initial parameter set $\vec{\theta}_{10}$. In Phase II, the parameter set $\vec{\theta}_{10}$ from Phase I was used as the seed for fine-tuning. This optimization continued using the classical Stochastic Gradient Descent (SGD)\cite{goodfellow2016deep} optimizer with a fixed learning rate of $1\times 10^{-3}$. 

To comparison with standard QAOA, we used the same SGD optimizer with random seed.

\section{Results}\label{result}
To quantify the efficacy of the meta-learning initialization approach, we evaluated the performance of the sequence model-QAOA frameworks on $90$ unseen Max-Cut graphs ($n=10$ to $n=13$). The overall performance was benchmarked against the standard random-seed QAOA baseline.

\begin{figure*}[t]
\centering
\includegraphics[width=\textwidth]{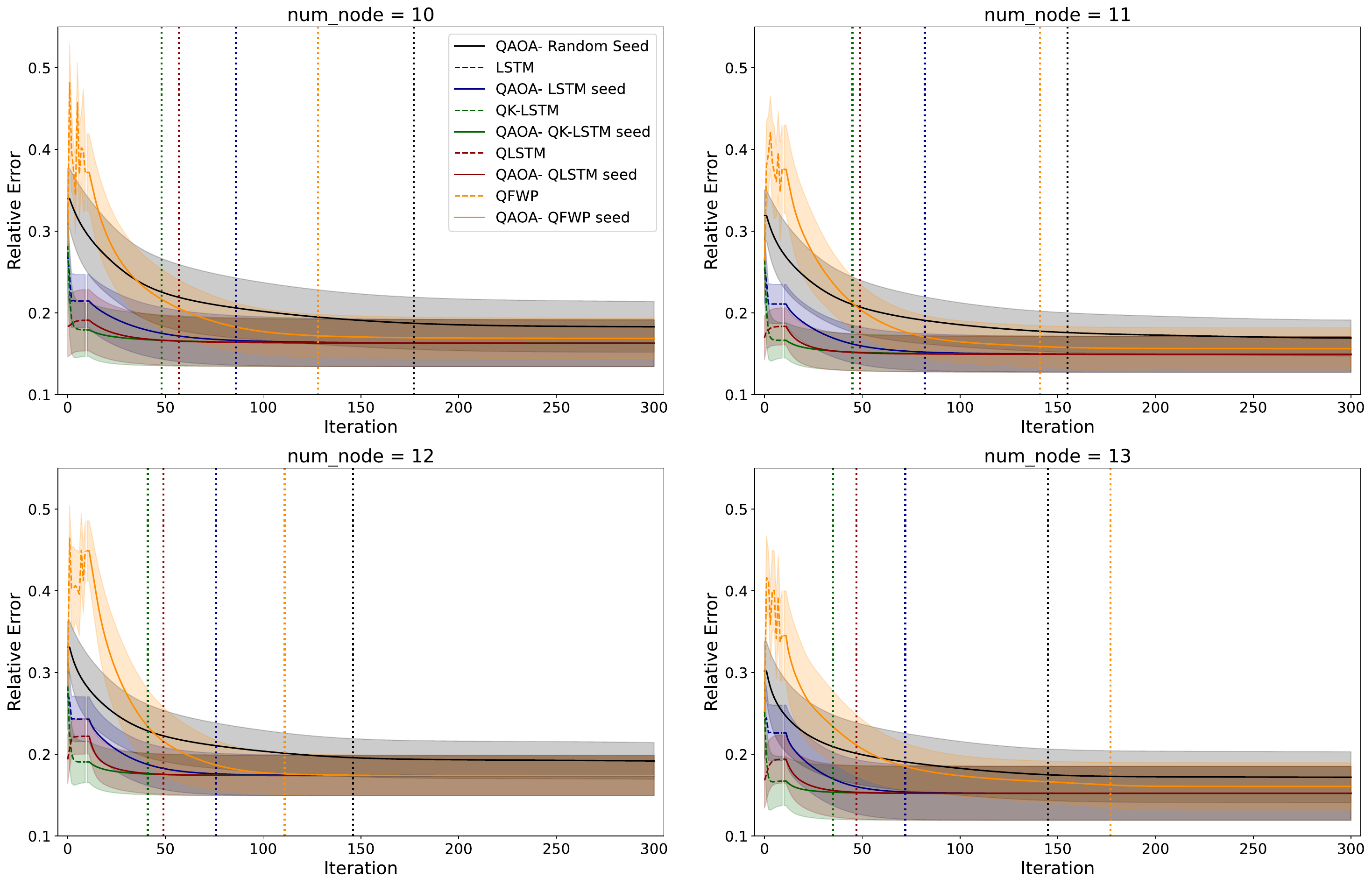}
\caption{Average relative errors over 300 iterations for sequence models on MaxCut problems with nodes $n=10$ to $13$. Dashed lines denote sequence model optimization; solid lines show post-seeding SGD optimization. Vertical lines indicate the average iteration to convergence (detailed in Table \ref{tab:iterations}). Error bars represent the $95\%$ confidence interval.}
\label{fig: relative error result}
\end{figure*}

Figure~\ref{fig: relative error result} displays the average relative error $\overline{f}_{\text{rel}}(j)$ as a function of the total optimization steps for Max-Cut instances with $n=10$ through $n=13$. The dashed curves graphically illustrate the optimization trajectory from the sequence model's recurrent steps (Phase I), while the subsequent solid lines represent the optimization path in Phase II, which was described in Section~\ref{sec:met}. The vertical dashed lines indicate the average iteration to convergence for each model, which is further detailed below. 

The relative error $f_{\text{rel}}(\vec{\theta})$ measures the gap between the expected cost value $f(\vec{\theta})$ and the globally optimal cost value $f_{\min}$, which was obtained via brute-force search.

 \allowdisplaybreaks
\vspace{-3pt}
\begin{align}
f_{\text{rel}}(\vec{\theta)}=\frac{f(\vec{\theta})-f_{\min}}{f_{\min}}
\end{align}
\vspace{-3pt}

The reported performance is based on the average relative error $\overline{f}_{\text{rel}(j)}$, which is defined as the average of the relative error at optimization step $j$ across all testing graph instances ($N_{\text{test}}$) for a given nodes $n$:

 \allowdisplaybreaks
\vspace{-3pt}
\begin{align}
\overline{f}_{\text{rel}}(j) = \frac{1}{N_{\text{test}}} \sum_{k=1}^{N_{\text{test}}} f_{\text{rel}}^{(k)}(j)
\end{align}
\vspace{-3pt}

Table \ref{tab:iterations} quantified the average number of iterations required for convergence for each sequence model-QAOA framework and the random-seed QAOA baseline across varying node sizes. The average iterations to convergence is defined as the total number of optimization steps $j$ (including the initial $T=10$ recurrent steps) required for the optimization trajectory to stabilize. Specifically, convergence is considered reached when the absolute step-to-step change in the average relative error falls below a tolerance of $\epsilon=10^{-4}$. 

 \allowdisplaybreaks
\vspace{-3pt}
\begin{align}
|\overline{f}_{\text{rel}}(j+1)-\overline{f}_{\text{rel}}(j)|\leq \epsilon
\end{align}
\vspace{-3pt}

\begin{table}[t]
    \caption{Average NUMBER of iterations to CONVERGENCE  (INCLUDE $T=10$ RECURRENT STEPS FOR SEQUENCE MODELS) FOR Node $10$ THROUGH $13$}
    \centering
    \label{tab:iterations}
    \resizebox{\columnwidth}{!}{%
    \begin{tabular}{l ccccc@{}}
        \toprule
        {\textbf{num\_node}} 
        & \textbf{QK-LSTM} & \textbf{LSTM} & \textbf{QLSTM} & \textbf{QFWP} & \textbf{Random seed} \\
        \midrule
        10 & \textbf{48} & 86 & 57 & 128 & 177 \\
        11 & \textbf{45} & 82 & 49 & 141 & 155\\
        12 & \textbf{41} & 76 & 49 & 111 & 146 \\
        13 & \textbf{35} & 72 & 47 & 177 & 145 \\
        \bottomrule
    \end{tabular}
    }
\end{table}

Table \ref{tab:approximation_ratio} details the approximation ratio \cite{chen2025L2L} (mean $\pm$ std) achieved at the end of the two phases described in Section \ref{sec:met}. The approximation ratio is defined as the ratio of the expected cut value $C(z)$ to the optimal cut value $C(z^*)$

\allowdisplaybreaks
\vspace{-3pt}
\begin{align}
\text{Approx. Ratio}(z)=\frac{C(z)}{C(z^*)}
\end{align}
\vspace{-3pt}

with the cut value $C(z)$ defined as: 

\allowdisplaybreaks
\vspace{-3pt}
\begin{align}
C(z)=\frac{1}{2}(\sum_{(i,j)\in E}w_{ij}-\sum_{(i,j)\in E} w_{ij}z_iz_j)
\end{align}
\vspace{-3pt}

This metric, bounded by $(0,1]$, is essential for assessing the quality of the final combinatorial solution.

\begin{table*}[t]
    \caption{APPROXIMATION RATIO (MEAN $\pm$ STD) ACHIEVED BY THESEQUENCE MOODEL (PHASE I, $T=10$) AND THE SUBSEQUENT QAOA WITH SEQUENCE MODEL SEEDING (PHASE II) AT ITERATION 10 FOR NODE 10 THROUGH 13. ALL VALUE ARE ROUNDED TO TWO DECIMALS.}
    \centering
    \label{tab:approximation_ratio}
    \begin{tabular*}{\textwidth}{@{\extracolsep{\fill}}l ccccccccc@{}}
        \toprule
        {\textbf{num\_node}} 
        & \multicolumn{4}{c}{\textbf{Phase I}} & \multicolumn{5}{c}{\textbf{Phase II}} \\\cmidrule(lr){2-5} \cmidrule(lr){6-10}
        & \textbf{QK-LSTM} & \textbf{LSTM} & \textbf{QLSTM} & \textbf{QFWP} & \textbf{QK-LSTM} & \textbf{LSTM} & \textbf{QLSTM} & \textbf{QFWP} & \textbf{Random seed} \\
        \midrule
        10 & \textbf{0.82 $\pm$ 0.03} & 0.79 $\pm$ 0.03 & 0.81 $\pm$ 0.04 & 0.63 $\pm$ 0.05 & \textbf{0.83 $\pm$ 0.03} & 0.80 $\pm$ 0.03 & 0.82 $\pm$ 0.03 & 0.70 $\pm$ 0.05 & 0.70 $\pm$ 0.05 \\
        11 & \textbf{0.83 $\pm$ 0.01} & 0.79 $\pm$ 0.02 & 0.82 $\pm$ 0.02 & 0.62 $\pm$ 0.05 &  \textbf{0.84 $\pm$ 0.01} & 0.81 $\pm$ 0.02 & 0.84 $\pm$ 0.02 & 0.70 $\pm$ 0.05 & 0.72 $\pm$ 0.04\\
        12 & \textbf{0.81 $\pm$ 0.03} & 0.76 $\pm$ 0.03 & 0.78 $\pm$ 0.02 & 0.55 $\pm$ 0.04 & \textbf{0.82 $\pm$ 0.03} & 0.78 $\pm$ 0.03 & 0.81 $\pm$ 0.03 & 0.66 $\pm$ 0.05 & 0.71 $\pm$ 0.04 \\
        13 & \textbf{0.83 $\pm$ 0.03} & 0.77 $\pm$ 0.03 & 0.81 $\pm$ 0.03 & 0.65 $\pm$ 0.05 & \textbf{0.84 $\pm$ 0.03} & 0.80 $\pm$ 0.03 & 0.83 $\pm$ 0.03 & 0.72 $\pm$ 0.05 & 0.75 $\pm$ 0.04\\
        \bottomrule
    \end{tabular*}
\end{table*}

\section{Discussion}\label{sec:dis}

As demonstrated in Figure \ref{fig: relative error result}, the QK-LSTM model achieved the highest quality approximate optimum of the QAOA parameters within the fixed 10 iteration compared to the other three sequence models and the standard QAOA with a random seed baseline. The results clearly show that the QK-LSTM framework provides the fastest and most stable convergence trajectory among all sequence models. 

This superior convergence translates directly into higher quality solutions. Table \ref{tab:approximation_ratio} reports the approximation ratios (mean $\pm$ std) at iteration 10. The QK-LSTM consistently exhibits the highest approximation ratio across various node sizes, underscoring its resilience and capacity to swiftly navigate the cost landscape to locate high quality solutions. While both QLSTM and LSTM also outperformed standard random initialization, their approximation ratios were notably lower than the quantum kernel enhanced model, confirming the specific advantage of the hybrid QK-LSTM architecture. 

\begin{figure}[t]
\centering
\includegraphics[width=\columnwidth]{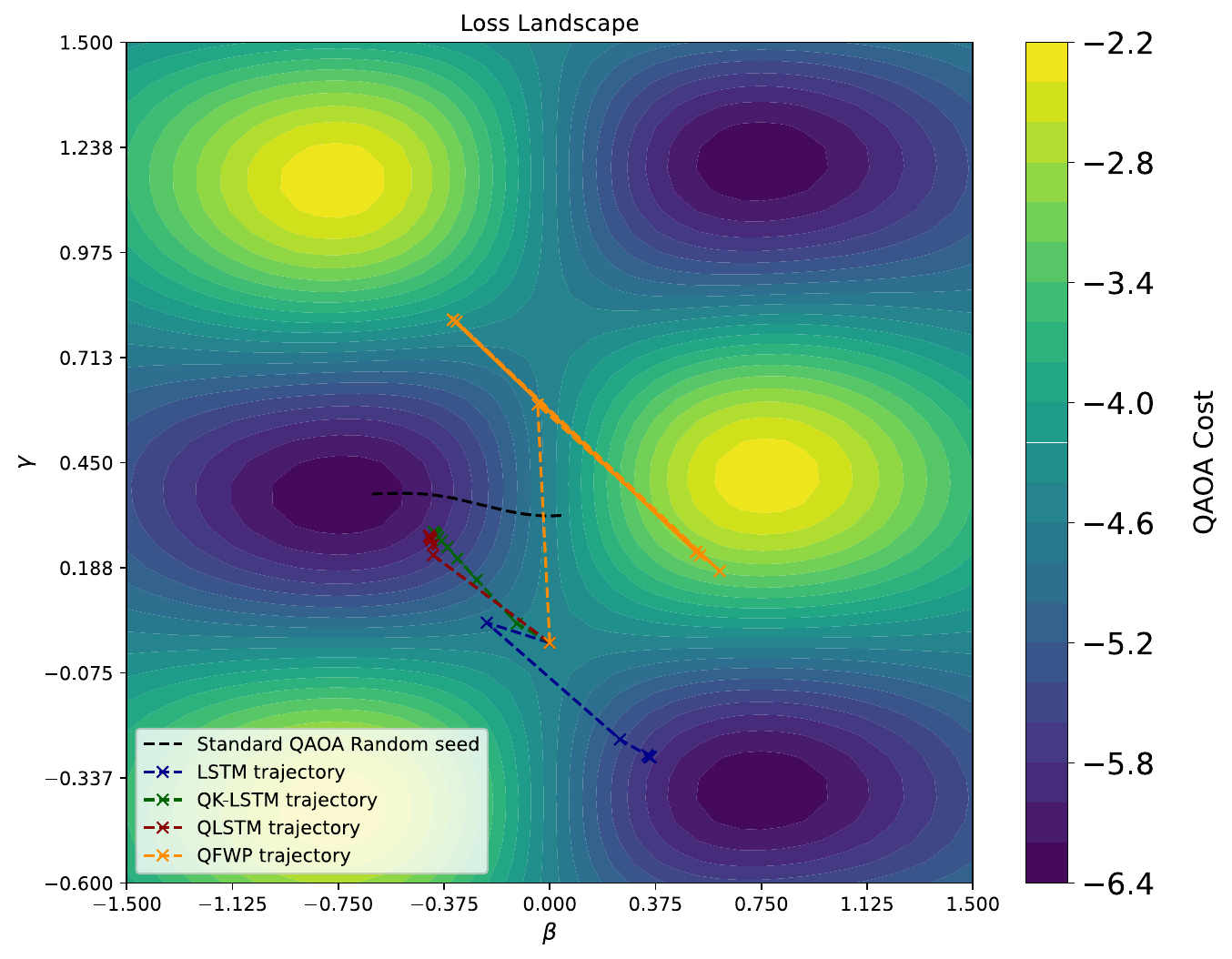}
\caption{Comparison of four sequence models and standard QAOA with random seed optimization trajectories in the single-layer QAOA parameter space, evaluated on a randomly generated graph with 10 nodes and edge probability $p=4/10$. All sequence models start from the same initial point $(0,0)$.}
\label{fig:landscape} 
\end{figure}

The success of the sequence models in this optimization task is rooted in their ability to quickly identify and propose initialization parameters that lie within a favorable basin of attraction on the QAOA cost landscape (see Fig. \ref{fig:landscape}). Figure \ref{fig:landscape} serves as an illustrative example of this mechanism, and consistent with this observation, parameter sets initialized by the QK-LSTM consistently led to the highest quality optima across all tested node sizes when compared to initializations derived from QLSTM, LSTM, QFWP, and the standard QAOA with a random seed. This finding validates the core hypothesis: the sequence models, particularly QK-LSTM, serve as highly effective meta-models capable of learning the optimal optimization trajectory from historical data.
 
The QK-LSTM model demonstrates significant strides in parameter efficiency by effectively incorporating the quantum kernel function within a classical LSTM architecture \cite{hsuLSTM2025quantum}. This integration not only leverages quantum feature spaces to capture intricate data dependencies but also achieves model compression with fewer trainable parameters than the LSTM model, all without sacrificing accuracy. This efficiency makes the QK-LSTM particularly suitable for deployment in resource-constrained environments. 

An interesting observation is that as the number of nodes increases, the average iteration count required for convergence of QK-LSTM, QLSTM, and LSTM decreases. This suggests that the meta-models are highly effective and potentially even better suited for the large-sclae problem sizes typical of challenge quantum optimization tasks, indicating that the predictive signal in the training data may become clearer with increasing graph complexity. 

It is important to note that the performance of our LSTM meta-model was lower than that observed in the prior literature  \cite{Guillaume2019lstmvqe}. This can be attributed to two main factors related to our experimental setup, which was fixed across all four models for a fair comparison: the use of a single, uniform learning rate for all models and our choice of a simpler meta-loss function, as opposed to the \textit{observed improvement} metric proposed in previous work \cite{Guillaume2019lstmvqe}.

\section{Conclusion and Future Work}\label{sec:con}
In this paper, we extended the application of meta-learning to QAOA optimization by proposing a novel comparison between four sequence models— QK-LSTM, QLSTM, LSTM, FWP, and standard QAOA—to solve the Max-Cut problem. 

Our numerical experiments highlight the QK-LSTM's superior performance. It demonstrated a remarkable ability to converge rapidly to near-optimal solutions, significantly outperforming the other three sequence models and the standard QAOA with random seed in terms of both iteration count and approximation ratios. We established that these models are used to quickly find a high quality global approximate optimum of the parameters, which then serves as a powerful initialization point for subsequent local search heuristic. This combined approach yielded high quality optima in the QAOA cost landscape, requiring substantially fewer quantum-classical optimization iterations than the traditional alternatives. Furthermore, the QK-LSTM's transfer-learning capability was shown to allow successful deployment on larger instances after training on smaller graphs, significantly reducing overall runtime and computational overhead.

In terms of possible extensions of this work, the meta-learning approach holds promise for application in several complex domains, including the Sherrington-Kirkpatrick Ising spin glass \cite{SK1975} and variational quantum eigensolver ansatz for preparing ground states of Hubbard models \cite{2018Hubbard} and molecular systems \cite{cheng2025lstmvqefc}. Future studies should also focus on testing the performance of these models under various hardware noise conditions. We envision that these future investigations will refine the quantum meta-learning paradigm, accelerating the practical and efficient deployment of quantum optimization algorithms in real-world realms such as materials science, logistics, and finance.

\section*{Code Availability}
All code and data used in this study are available at the following GitHub repository: \url{https://github.com/Astor-Hsu/QAOA-Meta-Learning}.

\section*{Acknowledgment}
The authors would like to thank the National Center for High-performance Computing of Taiwan for providing computational and storage resources. 

\bibliographystyle{ieeetr}
\bibliography{reference} 
\end{document}